\documentclass{elsart}
\usepackage{graphicx,amssymb}
\usepackage[english]{babel}

\newcommand{\kb}{\mathrm{k_B}}

\begin{document}
\begin{frontmatter}

\title{Temperature dependence of modified CNO nuclear reaction rates in dense stellar plasmas}

\collab{F.~Ferro$^a$, A.~Lavagno$^{a,b}$, P.~Quarati$^{a,c}$}
\address{$^a$Dipartimento di Fisica, Politecnico di Torino, I-10129 Torino, Italy} 
\address{$^b$INFN - Sezione di Torino, I-10125 Torino, Italy}
\address{$^c$INFN - Sezione di Cagliari, I-09042 Cagliari, Italy}

\begin{abstract}
We study the dependence of the CNO nuclear 
reaction rates on temperature, in the range of $10^7\div 10^8\,\mathrm{K}$, the typical range of temperature 
evolution from a Sun-like star towards a white dwarf. 
We show that the temperature dependence of the CNO nuclear 
reaction rates is strongly affected by the presence of non-extensive statistical effects in the dense stellar 
core. A very small deviation from the Maxwell-Boltzmann particle distribution implies a relevant enhancement of 
the CNO reaction rate and could explain the presence of heavier elements (e.g. Fe, Mg) in the final composition 
of a white dwarf core.  Such a behavior is consistent with the recent experimental upper limit to the fraction of 
energy that the Sun produces via the CNO fusion cycle.
\end{abstract}

\end{frontmatter}


\section{Introduction}
The classical theory of astrophysical reaction rates is based on the classical Boltzmann-Gibbs 
statistics~\cite{Rolfs,Clayton}. But, if we check for the applicability conditions inside a typical stellar 
plasma (e.g. inside the Sun or white dwarfs) one qualitatively obtains that~\cite{KLLQ}:
i) the collision time and the mean collisional time are of the same order; 
ii) the interaction is non-local and the random walks are not Markovian; 
iii) Boltzmann's hypothesis ({\em Stosszahlansatz}) is not rigorously satisfied; 
iv) the potential energy is almost equal to the thermal energy, i.e. the plasma parameter is 
not much smaller than one or of order unity~\cite{Plasma}.

It can be shown that a more suitable physical description relies on the non-extensive statistical 
mechanics~\cite{CKLLMQ,plb}. In the following, we will calculate non-extensive corrections to nuclear reactions rates 
of stellar plasma and to their chemical composition. 
During the evolution of a star from a Sun-like stage to the white dwarf stage, at high and very high 
temperature, the main nuclear burning cycle involves nuclei of intermediate weight (like carbon, nitrogen and 
oxygen). Therefore, we are primarily interested in studying the CNO cycle of reactions.

\section{The CNO reaction rates and luminosity}\label{luminosity}
The two most important nuclear cycles in stars, during their evolution, are the $pp$ chain and the CNO cycle, 
whose leading reactions are $p+p\rightarrow d+e^+ +\nu_e$ and $^{14}N+p\rightarrow^{15}O+\gamma$, respectively. 
The previous reactions account for the star's total energy production and luminosity~\cite{Rolfs}.

Given a reac\-tion bet\-ween two reac\-ting nuclei $i$ and $j$, the ap\-proxi\-ma\-te Max\-wellian (M) reaction rate, 
$r_{ij}^M$, is~\cite{Clayton}
\begin{equation}
r_{ij}^M\approx\frac{2^{5/2}}{3}{N_i N_j}\,\mu_{ij}^{-1/2}\frac{S}{(\kb T)^{1/2}}\;\tau^{1/2}\e^{\tau}\; , 
\label{nonres}
\end{equation}
where $N_i$ and $N_j$ are the particle densities, $\mu_{ij}$ is the reduced mass, $S$ is the astrophysical 
nuclear factor and $\tau\equiv 3E_0/(\kb T)$, $E_0$ being the most effective stellar energy (i.e. the energy at 
which the reactions are most likely to occur).

If we modify Eq.~\ref{nonres} with the non-extensive energy distribution function, the modified non-Maxwellian 
(NM) reaction rate, $r_{ij}^{NM}$, can be written as~\cite{Clayton paper}
\begin{equation}
r_{ij}^{NM}=r_{ij}^M\left(1+\frac{15}{4}\delta-\frac{7}{3}\delta\frac{E_0}{\kb 
T}\right)\exp\left(-\Delta_{ij}\right)\; ,\label{result}
\end{equation}
with
\begin{equation}
\Delta_{ij}=\frac{3E_0}{\kb T}\left[\left(1+\frac{5}{3}\delta\frac{\widetilde{E}_0}{\kb 
T}\right)\left(1+2\delta\frac{\widetilde{E}_0}{\kb T}\right)^{\hspace{-3pt}-2/3}-1\right]\; ,\label{delta}
\end{equation}
where the ``new'' most effective energy $\widetilde{E}_0$ is found to be the root of the equation 
$\widetilde{E}_0=E_0[(1+2\delta\widetilde{E}_0/(\kb T)]^{\hspace{-3pt}-2/3}$, and $\delta\equiv (1-q)/2$ is 
equivalent to the entropic index $q$. 

The deformed reaction rates $r_{ij}^{NM}$ of Eq.~\ref{result} give corrections to:
i) the luminosity fraction due to the $pp$ chain and to the CNO cycle (Section~\ref{luminosity});
ii) the chemical composition of the stellar plasma (Section~\ref{chemical}).

In the following, we choose the value $|\delta|=0.0045$ for the non-extensive deformation parameter, in order to 
work out a few reliable calculations. In Section~\ref{neutrino constraint} we will sketch an argument, based on 
recent results from solar neutrino experiments, that we have used to find out the previous value of $\delta$.

\begin{figure}[t]

\begin{center}
\includegraphics[width=.8\textwidth,height=.5\textwidth]{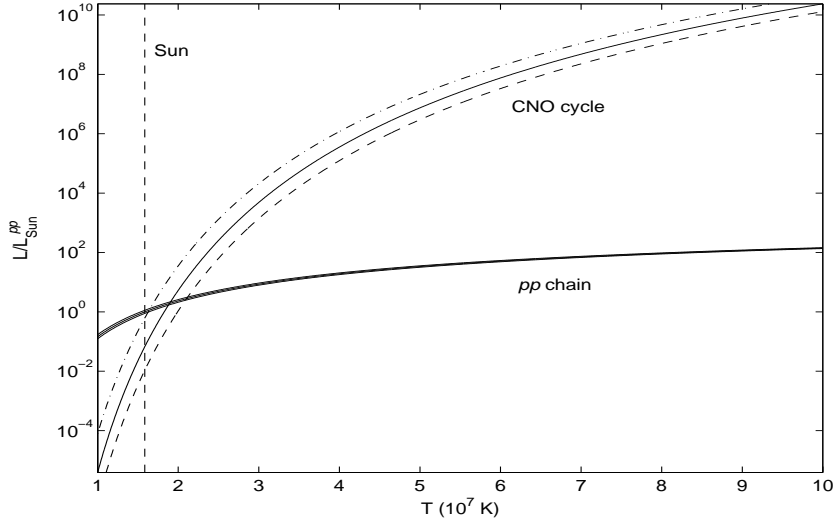}
\end{center}

\caption{Log-linear plot of dimensionless luminosity over temperature, for the $pp$ chain and the CNO cycle. 
(Dashed line, $\delta=+0.0045$, $q=0.991$; dash-dotted line, $\delta=-0.0045$, $q=1.009$). The vertical line 
shows the Sun's temperature. All curves are normalized with respect to the $pp$ luminosity inside the Sun.} 
\label{pp-CNO}

\end{figure}

In Fig.~\ref{pp-CNO}, the numerical curves of luminosity over temperature are plotted, along with non-extensive 
corrections.
It should be observed that: 
i) the luminosity yield of the $pp$ chain is slightly affected by the deformed statistics, with respect to the 
luminosity yield of the CNO cycle;
ii) the non-extensive CNO correction ranges from $37\%$ to more than $53\%$;
iii) above $T\approx 2\cdot 10^7\,{\rm K}$, the luminosity is mainly due to the CNO cycle only, thus 
confirming that CNO always plays a crucial role in the stellar evolution, when the star grows hotter toward the 
white dwarf stage.

\section{Chemical composition}\label{chemical}
The particle density of each nuclear species involved in the CNO cycle of reactions depends on the characteristic 
time constants of the nuclear reactions, that can be defined as
\begin{equation}
\tau\equiv\frac{1}{\langle\sigma v\rangle H}\; ,\label{time constant}
\end{equation}
where $H$ is the hydrogen density and $\langle\sigma v\rangle$ is the thermal average. To first approximation, 
the differential system of equations that describes the CNO cycle can be linearized, and it is cast in the 
straightforward formula
\begin{equation}
\frac{\d {\bf X}}{\d t}=\underline{\mathrm{A}}\,{\bf X}\; ,\label{CNOsystem}
\end{equation}
where ${\bf X}= (^{12}C\; ^{13}C\; ^{14}N\; ^{16}O\; ^{17}O)^T$  is the density vector, and 
$\underline{\mathrm{A}}$ is a $5\times 5$ real matrix whose elements are the time constants $\tau$ defined in 
Eq.~\ref{time constant}.

\begin{figure}[t]

\begin{center}
\includegraphics[width=.8\textwidth,height=.5\textwidth]{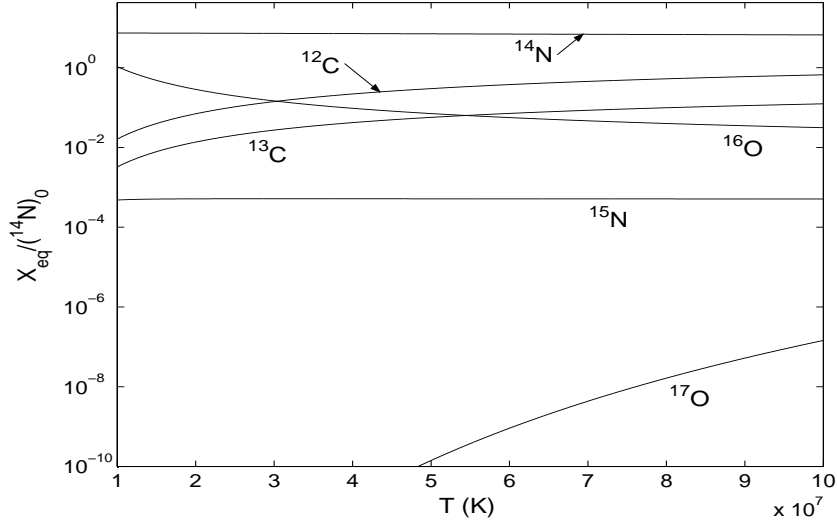}
\end{center}

\caption{Log-linear plot of dimensionless equilibrium concentrations of CNO nuclei over temperature. Classical 
statistics has been used. All curves are normalized with respect to the initial density $(^{14}N)_0$ inside the 
Sun (see Ref.~\cite{Clayton}).} \label{CNOnuclei figure}

\end{figure}

At time $t=0$, we choose the characteristic ratios~\cite{Clayton}
\begin{equation}
(^{12}C)_0\; :\; (^{14}N)_0\; :\; (^{16}O)_0=5.5\; :\; 1\; :\; 9.6\; ,\label{timezero}
\end{equation}
while the densities of all other chemical elements involved in the CNO cycle are supposed to be zero. The  
temperature dependence of a star's equilibrium chemical composition, i.e. the solution of Eqs.~\ref{CNOsystem} 
and~\ref{timezero}, is shown in Fig.~\ref{CNOnuclei figure}, considering classical statistical mechanics.

\begin{figure}[t]

\begin{center}
\includegraphics[width=.8\textwidth,height=.5\textwidth]{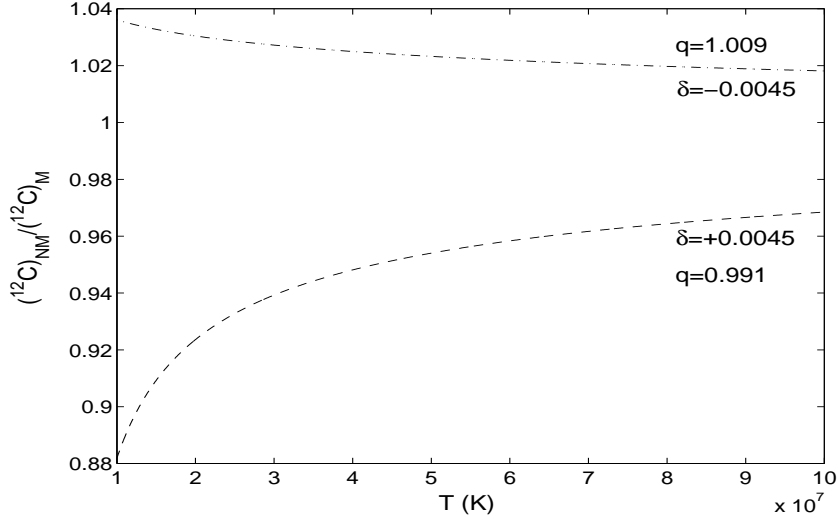}
\end{center}

\caption{Corrections to the concentration of $^{12}C$ over temperature, due to deformed statistics. (Dashed line, 
$\delta=+0.0045$, $q=0.991$; dash-dotted line, $\delta=-0.0045$, $q=1.009$).} \label{Carbon12 figure}

\end{figure}

If we now perform the previous calculations in the non-extensive statistics picture, each nuclear density might 
be corrected. For the sake of discussion, we have plotted in Fig.~\ref{Carbon12 figure} 
the corrective factor to the $^{12}C$ (that is far the 
most important medium-weighted nuclide) concentration as a function of the plasma 
temperature: the corrective factor is defined as the ratio $(^{12}C)_{NM}/(^{12}C)_M$, where $(^{12}C)_{NM}$ is 
the $^{12}C$ particle density with non-extensive statistics, while  $(^{12}C)_M$ is the classically calculated 
density, always with $|\delta|=0.0045$. 

We find that the deformed statistics affects the chemical composition of the plasma, and that the corrective 
factor is of order $10\%$, in the case of $^{12}C$ and $^{16}O$. On the contrary, if we consider the $^{14}N$ 
concentration, the corrective factor is about one order of magnitude smaller (and it can be neglected with 
respect to correction of $^{12}C$ and $^{16}O$).

\section{Neutrino constraint}\label{neutrino constraint}
Recent numerical calculations, performed on experimental data from the Sun collected over more than thirty 
years~\cite{Bahcall}, give an upper limit to the ratio between CNO luminosity and $pp$ luminosity in the Sun,
\begin{equation}
\left(\frac{L_{CNO}}{L_{\odot}}\right)\simeq 7.3\%\; ,\label{upper limit}
\end{equation}
from which the maximum allowable CNO neutrino flux is, classically,~\cite{BahcallExperimental,Bahcall-neutrino}
\begin{equation}
(\Phi_{CNO}^{max})_M\simeq 2.49\cdot 10^9 {\rm cm^{-2}s^{-1}}\; .\label{constraint}
\end{equation}

The maximum non-extensive neutrino flux, $(\Phi_{CNO}^{max})_{NM}$, can be calculated through the first order 
formula~\cite{KLLQ,neutrino}
\begin{equation}
(\Phi_{CNO}^{max})_{NM}\equiv (\Phi_{CNO}^{max})_M\,\exp(-\delta\beta)\; ,\label{neutrino formula}
\end{equation}
where $\beta\simeq 338.5$. From Eqs.~\ref{upper limit}, \ref{constraint} and~\ref{neutrino formula}, one can 
obtain a suitable value for the deformation parameter, namely,
\begin{equation}
|\delta|\simeq 0.0045\hspace{1cm} {\rm or}\hspace{1cm} 0.991\lesssim q\lesssim 1.009\; .\label{delta value}
\end{equation}

It should be observed that $|\delta|\simeq 0.0045$ is the maximum allowed value for the deformation parameter 
when considering the Sun's interior only. Therefore, the maximum value might be still greater when considering 
extreme physical conditions (e.g. high density plasmas in white dwarfs). 
This value agrees to the one derived from kinetic equations in presence of a random microfield distribution and 
given by the expression $\delta=12\, \alpha^4 \, \Gamma^2$, where $\alpha$ is a parameter related to ion-ion correlation 
and to the enforced Coulomb cross section \cite{plb}.

\section{Conclusions}
We have studied the temperature dependence of the CNO nuclear cycle in the range $10^7\div 10^8\,\mathrm{K}$. In 
this range of temperature, the CNO cycle plays a crucial role in the stellar evolution from a Sun-like star 
towards a white dwarf. We have found that a small deviation from the Maxwell-Boltzmann energy distribution 
function inside the stellar core implies relevant modifications on the nuclear reaction rates, luminosity and 
chemical composition of the plasma. \\
For $\delta<0$ (i.e. $q>1$), we obtain a remarkable increase of the CNO reaction rates with a more relevant 
contribution to the star luminosity with respect to the one obtained in the classical picture. Such a modification 
is consistent with the recent solar neutrino constraints that fix the deformation parameter to the value 
$|\delta|<0.0045$. The modified CNO reaction rates thus imply a faster evolution of stellar nuclear plasma 
towards heavier elements (like Mg and Fe) at high temperature. 
Such a behavior can be very relevant to understand the nature of matter in white dwarfs stars. In fact, 
a comparison of current observed mass-radius determinations with the 
theoretical curves seems to confirm that the composition of most white dwarfs is dominated by medium weight 
elements (carbon and oxygen). However, a small minority of white dwarfs do have relatively small radii indicating 
the presence of iron cores, which presents an intriguing puzzle from the point of view of stellar evolution~\cite{white dwarf}. 
A slight non-extensivity of the system could explain the presence of heavy elements in the final 
composition of white dwarf's core.


\end{document}